\documentstyle[twocolumn,aps,epsfig]{revtex}  
\newcommand{\msolar}{M$_{\odot}$}
\def\ensuremath#1{{\ifmmode #1 \else $#1$\fi}}

\newcommand{\nuc}[2]{\ensuremath{\mathrm {^{#2}#1}}}

\newcommand{\gk}{\ensuremath{\mathrm{\thinspace GK}}}
\begin{document}                
\title{Simulation of the Spherically Symmetric Stellar Core Collapse, 
Bounce, and Postbounce Evolution of a 13 M$_{\odot}$ Star with Boltzmann 
Neutrino Transport, and Its Implications for the Supernova Mechanism}
\author{
Anthony Mezzacappa$^{1}$, Matthias Liebend\"{o}rfer$^{1,2,3}$,
O. E. Bronson Messer$^{1,2,4}$,\\
W. Raphael Hix$^{1,2,4}$,
Friedrich-Karl Thielemann$^{1,3}$, Stephen W. Bruenn$^{5}$
}

\address{
$^{1}$ Physics Division, Oak Ridge National Laboratory, Oak Ridge, TN 
37831-6354 \\
$^{2}$ Department of Physics and Astronomy, University of Tennessee, Knoxville,
TN 37996-1200 \\
$^{3}$ Department of Physics and Astronomy, University of Basel,\\
Klingelbergstrasse 82, CH-4056 Basel Switzerland \\
$^{4}$ Joint Institute for Heavy Ion Research, Oak Ridge National 
Laboratory,\\
Oak Ridge, TN 37831-6374\\
$^{5}$ Department of Physics, Florida Atlantic University, Boca Raton,
FL 33431-0991
}
\maketitle
\pagebreak
\begin{abstract}                
With exact three-flavor Boltzmann neutrino transport, we simulate the 
stellar core collapse, bounce, and postbounce evolution of a 13 
\msolar\, star in spherical symmetry, the Newtonian limit, without 
invoking convection. In the absence of convection, prior spherically 
symmetric models, which implemented approximations to Boltzmann  
transport, failed to produce explosions. We are motivated to consider
exact transport to determine if these failures were due to the
transport approximations made and to answer remaining fundamental questions
in supernova theory. The model presented here is the first 
in a sequence of models beginning with different progenitors. In this 
model, a supernova explosion is not obtained. We discuss the ramifications of 
our results for the supernova mechanism.

\end{abstract}
\section{Supernova Paradigm}
Core collapse supernovae are among the most important phenomena
in astrophysics because of their energetics and nucleosynthesis.
Beginning with the first numerical simulations 
conducted by Colgate and White\cite{cw66}, three decades 
of supernova modeling have established a basic supernova paradigm.
The supernova shock wave---formed when the iron core of a massive 
star collapses gravitationally and rebounds as the core
matter exceeds nuclear densities---stalls 
in the iron core as a result of enervating losses to nuclear dissociation 
and neutrinos. The failure of this ``prompt'' supernova mechanism 
sets the stage for a ``delayed'' mechanism, whereby the shock is reenergized 
by the intense neutrino flux emerging from the neutrinospheres carrying 
off the binding energy of the 
proto-neutron star\cite{w85,bw85}. The heating is mediated primarily by the 
absorption of electron neutrinos and antineutrinos on the dissociation-liberated 
nucleons behind the shock. This past decade has also seen the emergence 
of multidimensional supernova models, which have investigated the 
role convection, rotation, and magnetic fields may play in the explosion
\cite{hbhfc94,bhf95,jm96,mcbbgsu98a,mcbbgsu98b,fh99,khowc99}.

Although a plausible framework is now in place, fundamental questions 
about the explosion mechanism remain: Is the neutrino heating sufficient, 
or are multidimensional effects such as convection and rotation necessary?
Can the basic supernova observable, explosion, be reproduced by 
detailed spherically symmetric models, or are multidimensional models  
required? 
Without a doubt, core collapse supernovae 
are not spherically symmetric. For example, neutron star kicks\cite{fbb98} 
and the polarization of 
supernova emitted light\cite{w99} 
cannot arise in spherical symmetry. Nonetheless, 
ascertaining the explosion mechanism
and understanding every explosion observable are two different 
goals. To achieve both, simulations in one, two, and three dimensions 
must be coordinated.

The neutrino energy deposition behind the shock depends sensitively 
not only on the neutrino luminosities but also on the neutrino spectra 
and angular distributions in the postshock region, necessitating exact
multigroup (multi-neutrino energy) Boltzmann neutrino transport. 
Ten percent variations in any of these quantities can make the 
difference between explosion and failure in supernova models\cite{jm96,bg93}. 
Past simulations have implemented increasingly sophisticated approximations 
to Boltzmann transport, the most sophisticated of which is 
multigroup flux-limited diffusion\cite{br93,wm93}. 
A generic feature of this approximation is that it 
underestimates the isotropy of the neutrino angular distributions in the 
heating region and, thus, the heating rate\cite{ja92,mmbg98}. It is important 
to note that, without invoking proto-neutron star (e.g., neutron finger) 
convection, simulations that implement multigroup flux-limited diffusion 
do not produce explosions\cite{br93,wm93}. Moreover, the existence and vigor of 
proto-neutron star convection is currently a matter of 
debate\cite{mcbbgsu98a,bd96,kjm96}.

Wilson\cite{w71} implemented an approximation to Boltzmann neutrino transport
by using order-of-magnitude parameterizations of the neutrino--matter
weak interactions. 
His models 
failed to produce explosions.
Core collapse simulations that implemented exact Boltzmann neutrino  
transport were completed by Mezzacappa 
and Bruenn\cite{mb93a,mb93c}. Following this work, we now present 
the findings of a core collapse, bounce, and
postbounce simulation.
Recognizing the need for more accurate time-dependent neutrino transport 
in supernova models, other groups have now developed
Boltzmann solvers\cite{yjs99,bypet99,rj00}.

\section {\bf Foundations}
We model the explosion of a 13 M$_{\odot}$ star, beginning with the
precollapse model of Nomoto and Hashimoto\cite{nh88}. The core collapse, bounce, 
and explosion were simulated with a new neutrino radiation hydrodynamics 
code for both Newtonian and general relativistic spherically symmetric flows: 
AGILE--BOLTZTRAN.
BOLTZTRAN is a three-flavor Boltzmann neutrino transport 
solver\cite{mb93b,mm99}, now extended to fully general
relativistic flows\cite{l00}. In this simulation, it is 
employed in the $O(v/c)$ limit with 6-point Gaussian quadrature to 
discretize the neutrino angular distributions and 12 energy groups
spanning the range from 5 to 300 MeV to 
discretize the neutrino spectra. 
AGILE is a conservative general relativistic hydrodynamics 
code\cite{l00,lt98}. 
Its adaptivity enables us to resolve and seamlessly follow the shock 
through the iron core into the outer stellar layers.

The equation of state of Lattimer and Swesty\cite{ls91} (LS
EOS) is employed to calculate the local thermodynamic state of
the matter in nuclear statistical equilibrium (NSE).
For matter initially in the silicon layer, the
temperatures are insufficient to achieve NSE.  In this
region, the radiation and electron components of the LS EOS are used, while an
ideal gas of \nuc{Si}{28} is assumed for the nuclear component. For typical 
hydrodynamic timesteps ($\sim .1$
millisecond), silicon burning occurs within a single timestep for T
$\sim 5 \gk$\cite{ht99}; therefore, when a fluid element exceeds a
temperature of 5 \gk\ in our simulation, the silicon is instantaneously burned, 
achieving NSE
and releasing thermal energy equal to the difference in nuclear binding
energy between \nuc{Si}{28} and the composition determined by the LS EOS.

We investigated the convergence of the net neutrino heating rate as the number 
of Gaussian quadrature points and the number of neutrino 
energy groups in our Boltzmann 
simulations were varied, as in Messer et al\cite{mmbg98}.
In the heating region, the 4- and 6-point rates, the 6- and 8-point rates, 
and the 12- and 20-group rates differed by at most 5 percent, 3 percent, 
and 3 percent, respectively. 
Moreover, during the course of the important first 300 ms of 
our simulation, the maximum variation in
the total energy is $\sim 3\times 10^{49}$ erg, which is a few percent 
of the total energy, and the total lepton number 
is conserved to within a fraction of a percent.
Note that the numerical uncertainty in the net heating rate 
(which is at most 3 percent in our model) is no
larger than the uncertainty in the total energy. Therefore,
any further numerical convergence in the computation of this 
rate would be meaningless.

\begin{figure}
\begin{center}
\epsfig{file=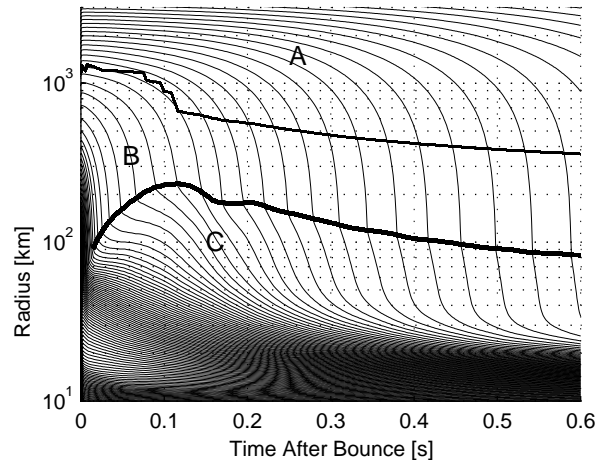}
\end{center}
\caption{Radial trajectories of equal mass shells in the iron core and
silicon layer. We also trace the shock, nuclear burning, and dissociation 
fronts, which carve out three regions in the $(r,t)$ plane. 
A: Silicon. 
B: Iron produced by infall compression and heating. 
C: Free nucleons and alpha particles. 
}
\label{fig1}
\end{figure}

\section{\bf Stellar Core Collapse, Bounce, and Postbounce Evolution}
Figure 1 shows the radius-versus-time trajectories of equal mass 
shells (0.01M$_{\odot}$) in the stellar iron core and silicon layer
during the first 600 ms of postbounce evolution. It also shows the
shock and nuclear burning front trajectories.
At 110 ms after bounce, the shock stalls at a radius of
230 km and then recedes for the duration of the simulation, and no
explosion is launched. The 
shock and burning fronts divide the stellar core and silicon 
layer into three regions: 
A: Silicon. 
B: Iron produced by infall compression and heating. 
C: Free nucleons and alpha particles. 

\begin{figure}
\begin{center}
\epsfig{file=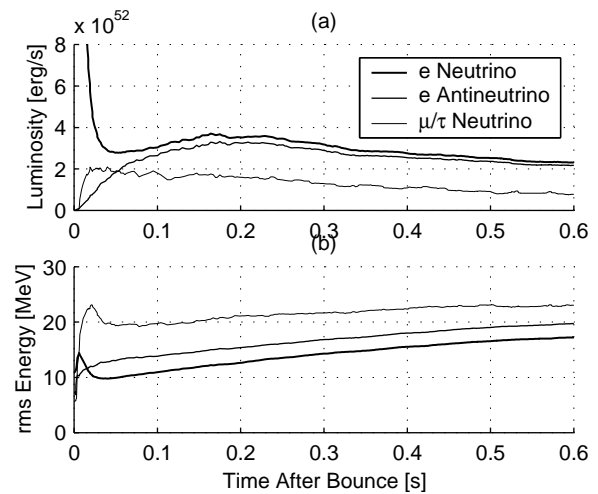}
\end{center}
\caption{Three-flavor neutrino luminosities and rms energies at a 
radius of 500 km in the core as a function of time during our simulation.}
\label{fig2}
\end{figure}

In Figure 2, we plot the neutrino luminosities and rms energies at
500 km in the stellar core as a function of time. The electron neutrino 
luminosity decreases from its early ``burst'' value of $3.5\times 10^{53}$ 
erg/s as we enter the postbounce accretion phase. (The early electron 
neutrino burst occurs as the shock passes the electron neutrinosphere 
in the core, liberating the trapped neutrinos behind it that are produced 
by electron capture during stellar core collapse.) In the accretion 
phase, the electron neutrino 
luminosity reaches a maximum of $3.6\times 10^{52}$ erg/s 
and then decreases slowly as the mass accretion rate decreases. The electron 
antineutrino and muon/tau neutrino and antineutrino luminosities rise after 
a hot, deleptonized ``mantle'' forms beneath the shock (this is the region
above the cold, degenerate, unshocked core, and includes the neutrinospheres).
In this mantle, electron antineutrinos are produced by positron capture and all 
three flavors of neutrinos and antineutrinos are produced by electron--positron 
annihilation. The electron antineutrino luminosity reaches a maximum 
of $3.3\times 10^{52}$ erg/s and exhibits the same subsequent 
decrease with decreasing mass accretion rate. The muon/tau neutrino and 
antineutrino luminosities on the other hand reach values of only $2.0\times 
10^{52}$ erg/s, and thereafter decrease with time.
For all three flavors, the rms energies increase with time, owing
to infall into an increasingly deep gravitational well. Relative to
the electron neutrino and antineutrino rms energies, the muon/tau neutrino and 
antineutrino rms energies are larger, with values between 20--25 MeV: 
the muon/tau neutrinos and antineutrinos 
interact only via neutral currents and therefore decouple deeper 
in the core at higher densities. The electron neutrino rms energies 
lie in the range between 10--20 MeV. Relative to the electron 
neutrino rms energies, the electron antineutrino rms energies
are slightly larger because the electron antineutrinos decouple 
at slightly higher densities: the core 
material is neutron rich; therefore, electron antineutrino absorption 
on protons is reduced relative to electron neutrino absorption on neutrons.

In Figure 3, we plot the mass density, entropy per baryon (in units
of $k_{\rm B}$), electron fraction, and velocity as a function of radius 
for various time slices in our simulation. In the velocity profiles,
the initial outward propagation of the shock is evident, as is the 
subsequent decrease in radius at later times. Note the increasing 
infall velocities below the shock as it recedes, reaching values 
of -6000 km/s at 600 ms after bounce. Despite the decreasing
shock radius and failed explosion, there remains a heating region 
behind the shock, where the entropies continue to increase. This
is evident in the entropy profiles. Nonetheless, conditions remain 
such that an explosion does not occur in this model in the first 
second.

\begin{figure}
\begin{center}
\epsfig{file=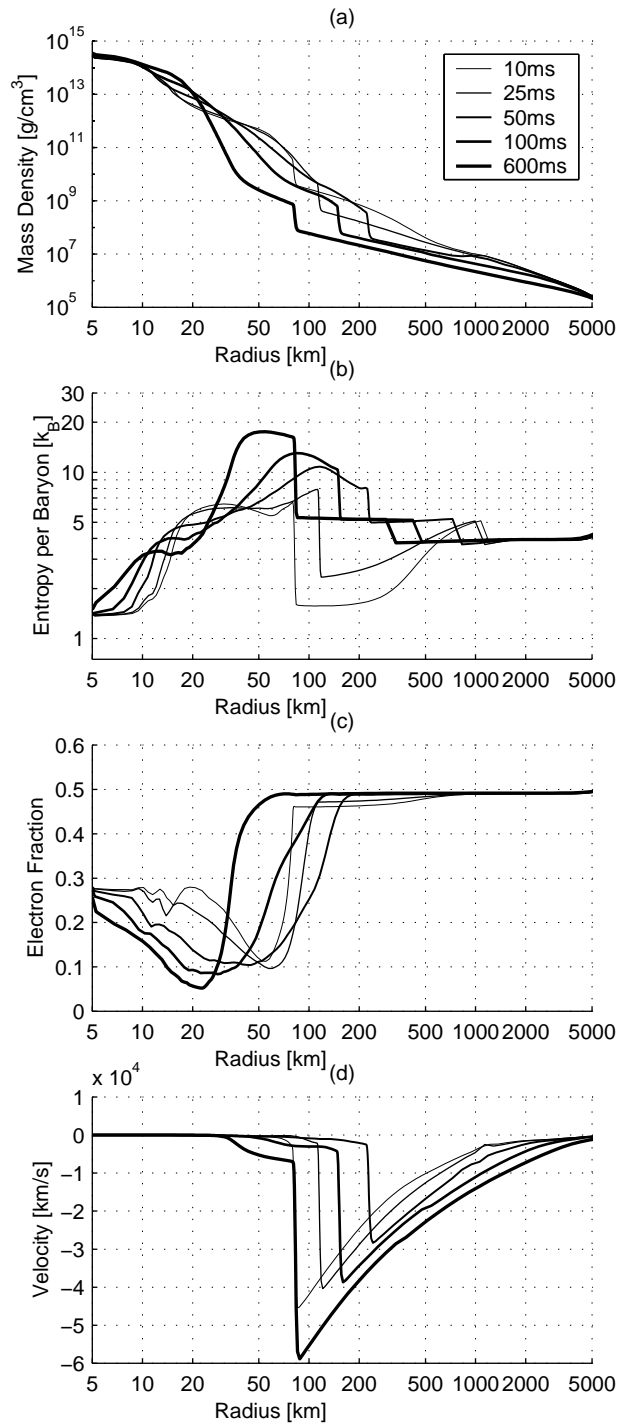}
\end{center}
\caption{Core baryon density, entropy per baryon, electron fraction,
and velocity profiles at select times during our simulation.}
\label{fig3}
\end{figure}

\section{\bf Outlook}
We have presented results from the simulation of the stellar core collapse, 
bounce, and the first 600 ms of postbounce evolution of a 13 M$_{\odot}$ 
progenitor. Spherical symmetry was assumed, $O(v/c)$ Boltzmann neutrino 
transport was implemented, and gravity was Newtonian. No explosion was 
obtained. In light of our implementation of Boltzmann transport, if we 
do not obtain explosions in models initiated from other progenitors (see 
also Rampp and Janka\cite{rj00}), it would indicate that improvements in 
our initial conditions (precollapse models) and/or input physics are needed, 
and/or that the inclusion of multidimensional effects such as convection, 
rotation, and magnetic fields are required ingredients in the recipe for 
explosion. In the past, it was not clear whether failures to produce explosions 
in spherically symmetric models were the result of transport approximations 
or the neglect of an important physical effect. 
We will report on the general relativistic case\cite{l00,lmtmhb00} and other models 
in subsequent papers.

Potential improvements in our initial conditions and input physics
include: improvements
in precollapse models\cite{ba98,unn99,hlw00,hlmw00}; the use of ensembles of
nuclei in the stellar core rather than a single representative nucleus;
determining the electron capture rates on this ensemble
with detailed shell model
computations\cite{lm00}; the inclusion of nucleon correlations in the
high-density neutrino opacities\cite{bs98,rplp99}; and the inclusion
of new neutrino emissivities in dense matter\cite{tbh00}. These improvements
all have the potential
to quantitatively, if not qualitatively, change the details of our simulations.
Thus, it is important to note that our conclusions are drawn
considering the initial conditions and input physics used.

\section*{Acknowledgments}
A.M. is supported at the Oak Ridge National Laboratory, managed by
UT-Battelle, LLC, for the U.S. Department of Energy under contract
DE-AC05-00OR22725. 
M.L. is supported by the National Science Foundation under contract 
AST-9877130 and, formerly, was supported by the Swiss National 
Science Foundation under contract 2000-53798.98. 
O.E.B.M. is supported by funds from the Joint Institute for Heavy Ion 
Research and a Department of Energy PECASE award. 
W.R.H. is supported by NASA under contract NAG5-8405 and by funds 
from the Joint Institute for Heavy Ion Research.
F.-K.T. is supported in part by the Swiss National Science Foundation 
under contract 2000-53798.98 and as a Visiting Distinguished Scientist
at the Oak Ridge National Laboratory. 
S.W.B. is supported by the NSF under contract 96-18423 and by NASA under
contract NAG5-3903.
Our simulations were carried out on the ORNL Physics Division Cray J90 
and the National Energy Research Supercomputer Center Cray J90 and
SV-1. We thank the referees for their important questions and 
comments.

\end{document}